%% file: scr.tex
\documentclass[12pt]{article}
\usepackage{amssymb,epsfig,bm,color,slashed,mathrsfs}
\usepackage[german,english]{babel}
\newcommand{\pp}{{\rm pp}}
\newcommand{\ph}{{\rm ph}}

\newcommand{\be}{\begin{equation}}
\newcommand{\ee}{\end{equation}}
\newcommand{\bea}{\begin{eqnarray}}
\newcommand{\eea}{\end{eqnarray}}

\newcommand{\vecnabla}{{\bm \nabla}}

\newcommand{\vecq}{{\bm q}}
\newcommand{\vecp}{{\bm p}}

\newcommand{\vecg}{{\bm g}}

\newcommand{\vecv}{\bm v}

\newcommand{\ie}{{\it i.e.}}

\definecolor{red}{rgb}{0.8,0,0}
\definecolor{violet}{rgb}{0.4,0,0.4}
\definecolor{green}{rgb}{0,0.5,0.0}
\definecolor{GREEN}{rgb}{0,0.5,0.0}
\definecolor{navy}{rgb}{0.0,0.0,0.6}
\definecolor{orange}{rgb}{0.8,0.2,0.0}
\definecolor{blue}{rgb}{0.3,0.0,0.8}

\input econfmacros-csqcd4.tex
\def\Title#1{\begin{center} {\Large {\bf #1} } \end{center}}

\begin{document}

\Title{Spin diffusive modes and thermal transport in neutron-star crusts}

\bigskip\bigskip

\begin{raggedright}
{\it 
Armen Sedrakian$^{1}$~and John W. Clark$^{2}$\\
\bigskip
$^{1}$Institute for Theoretical Physics, J.~W.~Goethe University,  D-60438  Frankfurt am Main, Germany\\
\bigskip
$^{2}$Department of Physics and McDonnell Center for the Space
Sciences, Washington University, St.~Louis, Missouri 63130, USA\\
}
\end{raggedright}

\section{Introduction}
\label{sec:1}

In this contribution we review a method for deriving collective
modes of pair-correlated neutron matter as found in the inner crust
of a neutron star.  We discuss two classes of modes lying either 
above or below the pair-breaking continuum limit $2\Delta$ and having 
energy spectra of the form $\varepsilon = \omega_0 + \alpha q^2$, 
where ${\vecq}$ is the wave vector of the mode and $\omega_0 = 2\Delta$ 
is a threshold frequency determined by the pairing gap $\Delta$. 
One class of modes, corresponding to density oscillations, arises 
in the spectrum for small $q$ when $\alpha < 0$.  These modes can be 
associated with undamped particle-hole bound pairs -- {\it excitons} 
-- existing in superfluid neutron matter.  We also discuss the 
modes with $\alpha >0$. These are {\it spin diffusive modes} that 
arise in neutron-star crusts due to spin perturbations and are 
located above the pair-breaking threshold. In contrast to excitons,
these modes are damped.  As an application we compute the thermal
conductivity due to the spin diffusive modes existing in the crust 
of a neutron star.  The general formalism that we introduce below 
is relevant for a number of fermionic ensembles experiencing
attractive interactions, which become superfluid or superconducting 
below a critical temperature $T_c$. Examples are ultra-cold 
atomic vapors, nuclear and neutron matter, and deconfined quark 
matter~(for review articles see Ref.~\cite{SedrakianClarkAlford}).

Microscopic description of the low-energy dynamics of superfluid
systems can be formulated within the propagator formalism, where
the central quantity is the response function. Several types of response
functions may be considered, depending on the problem of interest.
For neutral fermionic systems, the central quantities are the density
and spin responses. For charged superconductors the charge current
response has been especially important historically, since it describes
the experimentally observed Meissner effect. In the present work we
focus instead on the pair-breaking response of a neutral,
non-relativistic superfluid exhibiting $S$-wave pairing, a
response which is dominant at temperatures slightly below the 
critical temperature of the superfluid phase transition.

Collective modes of neutron-star inner crusts, which are composed
of an ionic lattice, a charge-neutralizing electron background, and 
a superfluid neutron fluid, have been studied from a number of 
perspectives.  The electron-ion system oscillates at the plasma 
frequency in the absence of the neutron fluid~\cite{FlowersItoh1976}. 
However, coupling of the lattice phonons to the neutron fluid 
renormalizes the spectrum of lattice oscillations \cite{Sedrakian1996} 
and may lead to instabilities~\cite{Kobyakov:2013pza}.  The 
low-lying oscillatory modes with spectra of acoustic type,
notably the Anderson-Bogolyubov mode and the lattice phonon mode, 
tend to mix since they cover the same energy range.  Their 
properties and mixing have been investigated in
Refs.~\cite{Pethick:2010zf,Cirigliano:2011tj,Chamel:2012ix,Keller:2012yc,Martin:2014jja}. 
In a first approximation, the pair-breaking modes can be treated 
separately, as they arise near the pair-breaking threshold $2\Delta$.  
Before discussing concrete applications to neutron-star crusts, we 
first outline the general formalism for deriving these modes,
following closely Ref.~\cite{Keller:2012yc}.

\section{Formalism}

The techniques we employ were first developed by Abrikosov and
Gor'kov (AG) to study the Meissner effect in the electrodynamics of dirty
superconductors~\cite{Abrikosov:1962}.  In
this theory the electromagnetic response of the system is expressed
in terms of Matsubara imaginary-time propagators, assuming a contact
pairing interaction. From the formal standpoint, the AG theory can 
be viewed as a reformulation of the theories of collective excitations 
in superconductors developed by Bogolyubov, Anderson, and subsequently 
others using the second-quantized operator formalism.  The ideas of 
Landau Fermi-liquid theory (which is applicable to fermionic systems 
with arbitrarily strong interactions) were extended to superfluid systems 
by Larkin and Migdal and by Leggett ~\cite{Migdal:1967,Leggett:1966zz}. They
reproduced the basic results known at the time, going on to derive
the so-called Fermi-liquid corrections that arise in strong-coupling
theories. In contrast to the AG treatment, the latter approaches
distinguish particle-hole ($\ph$) and particle-particle ($\pp$)
interactions having different strength/or sign, in the spirit of 
the Fermi-liquid theory.  There is renewed interest in studying the 
response of superfluid/superconducting nuclear matter, in part 
with the objective of understanding the collective modes of such 
systems (see Ref.~\cite{Keller:2012yc} and works cited therein).

In contributing to this trend, we begin by considering $S$-wave 
paired neutron matter and write the interactions in the 
particle-particle and particle-hole channel in the 
forms~\cite{Migdal:1967}
\bea \hat
f^{\pp}_{\alpha\beta\gamma\delta} &\simeq &
f^D_{\pp}(i\sigma_2)_{\alpha\beta}(i\sigma_2)_{\gamma\delta} +
f^S_{\pp}({i\sigma_2\bm{\sigma}})_{\alpha\beta}\cdot
({\bm{\sigma}}i\sigma_2)_{\gamma\delta},
\\
\hat f^{\ph}_{\alpha\beta\gamma\delta} &\simeq
&f^D_{\ph}\delta_{\alpha\beta}\delta_{\gamma\delta} +
f^S_{\ph}{\bm{\sigma}}_{\alpha\beta}\cdot{\bm{\sigma}}_{\gamma\delta}.
\eea 
Here $f^D$ and $f^S$ are constants that parametrize the two-body
interaction in the density and spin channels, respectively,
the superscripts $\pp$ and $\ph$ refer to the particle-particle and
particle-hole channels, while the quantities ${\bm{\sigma}}_{\alpha\beta}$
are Pauli matrices in the spin space.

\begin{figure*}[tb]
\begin{center}
\includegraphics[height=1.3cm,width=12.cm]{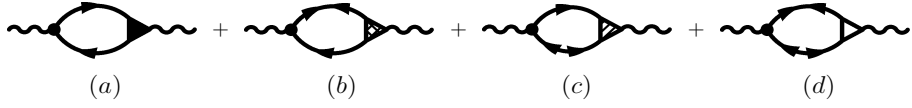}
\end{center}
\caption[] {Sum of polarization tensors contributing to the
  response function of a superfluid. The diagrams $b$, $c$, and $d$ are
  specific to superfluid systems and vanish in the unpaired state. }
\label{fig3}
\end{figure*}
The low-lying excitations of neutron matter can be obtained by
expanding the relevant response functions in small parameters of the
theory, for example $q/k_F$, where $k_F$ is the Fermi
wave number and $q$ is the magnitude of the momentum transfer.
If we are interested in radiation processes, i.e., in the time-like
kinematical domain, the quantity $q v_F/\omega$ is small, where
$v_F$ is the Fermi velocity and $\omega$ is the energy transfer
(typically of the order of the temperature).

As there are four distinct propagators in the theory of superfluid
systems (the so-called Nambu-Gor'kov structure), the response function
is given by four different contributions, shown as Feynman diagrams
in Fig.~\ref{fig3}.  We adopt the standard notation of the diagram
technique for superfluid systems, following Ref.~\cite{Abrikosov:1962}.
The shaded (full) vertices appearing in the polarization loops can be
computed, in turn, by resummation of infinite loop diagrams, which
account for the modifications due to the interactions in the medium
(see for example Ref.~\cite{Keller:2012yc}).  To leading order in a
small-momentum expansion, the temporal ($00$) and spatial ($jj$)
components of density ($D$) and spin ($S$) response functions may
be written as
%\cite{Kolomeitsev:2008mc,Leinson:2009mq,Kolomeitsev:2010hr,Sedrakian:2012ha}
%--------------------------------------------------------------------
\bea
\label{eq:pi_density}
\Pi_D^{00}
&=& -\frac{4 q^4 v_F^4}{45\omega^4}\,\mathcal{G},\quad 
\Pi_D^{jj}
 = -\frac{2 q^2 v_F^4}{9\omega^2}\,\mathcal{G},\\
\label{eq:pi_spincurrent}
\Pi_S^{00}&=&-v_F^2\,\mathcal{G},\quad \quad 
\Pi_S^{jj}
 = -\frac{q^2 v_F^2}{\omega^2}\,\mathcal{G} ,
\eea
%--------------------------------------------------------------------
where 
%--------------------------------------------------------------------
\bea
\label{eq:Gthermal}
\mathcal{G}(\vecv,\omega,\vecq)
=  \Delta^2\!\!\!\int\limits_{-\infty}^{+\infty}\!d\xi_p\!
\Biggl[\frac{\epsilon_+-\epsilon_-}{\epsilon_+\epsilon_-}
\frac{f(\epsilon_-)-f(\epsilon_+)}
{\omega^2-(\epsilon_+-\epsilon_-)^2+i\eta
  }
-\frac{\epsilon_++\epsilon_-}{\epsilon_+\epsilon_-}
  \frac{1-f(\epsilon_-)-f(\epsilon_+)}{\omega^2-(\epsilon_++\epsilon_-)^2+i\eta}
  \Biggr].\nonumber
\eea
%--------------------------------------------------------------------
Here $f(x) = \{\exp[(x-\mu)/T]+1\}^{-1}$ is the Fermi function, $\mu$ being
the chemical potential.  We define $\epsilon_\pm=\sqrt{\xi_\pm^2+\Delta^2} $
and $\xi_\pm =\xi_p\pm{\vecq\vecv}/{2}$, where $\xi_p$ is the
spectrum in the unpaired state, $\Delta$ is the pairing gap,
$\vecv$ is the particle velocity, and $\vecq$ is the momentum transfer.
In concentrating on the pair-breaking contribution, we need only keep
the term proportional to $ 1-f(\epsilon_-)-f(\epsilon_+)$.

\begin{figure}[t]
\begin{center}
\includegraphics[width=7.0cm,height=5.0cm,angle=0]{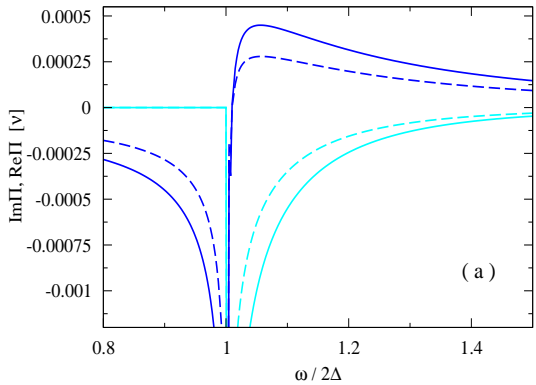}
\caption[] {Comparison between numerical (solid lines) and perturbative 
  (dashed lines) results for the density response function $\Pi^D_{00}$, 
  in units of the density of states $\nu$~\cite{Keller:2012yc}.  The 
  imaginary part of the response function is shown by a heavy (blue) 
  line, the real part by a light (cyan) line. The energy transfer $\omega$ 
  is in units of $\omega_0$.  The temperature is $0.5\,T_c$ and the pairing
  gap is $\Delta=1.0~\textrm{MeV}$.  The ratio of momentum transfer and
  Fermi momentum is kept fixed at $q/k_F=0.01$, and the Fermi momentum
  is $k_F=1.0\,\textrm{fm}^{-1}$ (which corresponds to a density
  $n=0.22n_0$ in terms of the nuclear saturation density $n_0$.) 
 }
\label{fig:reimpi}
\end{center}
\end{figure}

Figure~\ref{fig:reimpi} illustrates the dependence on the transferred
energy $\omega$, of the real and imaginary parts of the density response
function of neutron matter, for fixed three-momentum transfer.  The
zero-temperature gap is related to $T_c$ via $T_c =\Delta(0)/1.76$. 
The Landau parameter is set to $f_{\ph}^D = -0.5$ for density perturbations,
while the frequency and momentum transfer are normalized to the threshold
frequency $\omega_0 = 2\Delta(T)$.  The response function in the negative
energy range can be obtained from the relations ${\rm Re}\Pi^{D}(-\omega)
= {\rm Re}\Pi^{D}(\omega)$ and ${\rm Im}\Pi^{D}(-\omega) =
-{\rm Im}\Pi^{D}(\omega)$. There exists a threshold in the response
function because an energy $\omega \ge \omega_0$ is needed to break
a Cooper pair, i.e., the energy transfer must exceed the binding energy
of the pair. For energies above the threshold ($\omega > \omega_0$), a
non-zero imaginary part implies that the collective excitations have
finite life-time.  They are not perfect quasiparticles.

%-------------------------------------------------------------------

\section{Numerical results and collective modes}
\label{sec:5}

The numerical results reported here were obtained in Ref.~\cite{Keller:2012yc}.
Once the response function is determined, we are in a position to write
down the associated spectral function, which is defined by
$R(\omega,\vecq) =-2{\rm Im}\Pi(\omega,\vecq)$ and can be cast as 
%--------------------------------------------------------------
\bea
R(\omega,\vecq)= -\frac{2 (v_{\ph}^D)^{-2}{\rm Im} {P}(\omega,\vecq)}
{\left[(v_{\ph}^D)^{-1} - {\rm Re} {P}(\omega,\vecq) \right]^2 + {\rm
    Im}{P}(\omega,\vecq)^2}, \eea where
${P} (\omega,\vecq)\equiv
{\mathcal{Q}^+(\omega,\vecq)}/{\mathcal{C}(\omega,\vecq)} $ and 
%--------------------------------------------------------------
\bea 
\mathcal{Q}^{\pm}(\omega,\vecq)
= \mathcal{A}^{\pm}(\omega,\vecq)\mathcal{C}(\omega,\vecq)
   -\mathcal{B}(\omega,\vecq)\mathcal{D}^{\pm}(\omega,\vecq).
\eea
The elementary loops are given by~\cite{Migdal:1967,Leggett:1966zz}
%-------------------------------------------------------------------
\bea
\label{eq:LoopA}
\mathcal{A}^{\pm}
&=&\nu\int\!\frac{d{{\Omega}}}{4\pi}\,
   \Biggl\{-\frac{1\pm\hat{\mathcal{P}}}{2}\,\mathcal{G}(\vecv,\omega,\vecq)
    +\frac{\vecq\vecv}{\omega-\vecq\vecv}
   \,\Big[\mathcal{G}(\vecv,\vecq\vecv,\vecq)-\mathcal{G}(\vecv,\omega,\vecq)\Big]\Biggr\},
\\
\label{eq:LoopB}
\mathcal{B}
&=&-\nu\int\!\frac{d{{\Omega}}}{4\pi}\,
   \frac{\omega+\vecq\vecv}{2\Delta}\,\mathcal{G}(\vecv,\omega,\vecq),\\
\label{eq:LoopC}
\mathcal{C}
&=&\nu\int\!\frac{d{{\Omega}}}{4\pi}
   \frac{\omega^2-(\vecq\vecv)^2}{4\Delta^2}\,\mathcal{G}(\vecv,\omega,\vecq),\\
\label{eq:LoopD}
\mathcal{D}^\pm &=&\nu\int\!\frac{d{{\Omega}}}{4\pi}\,
\left[\frac{\omega+\vecq\vecv}{4\Delta}
  +\frac{\omega-\vecq\vecv}{4\Delta}\hat{\mathcal{P}}\right]\,
\mathcal{G}(\vecv,\omega,\vecq),
\eea 
%-----------------------------------------------------------------------
where the operator $\hat{\mathcal{P}}$ is equal to $+1$ for vertices 
which are even under the time reversal operation and $-1$ for
vertices odd under this transformation, while $\nu$ is the density
of states. 

For small  imaginary parts the spectral function can be approximated 
as 
\bea
\label{S_SMALL} 
R(\omega,\vecq) = 
2\pi Z(\vecq) \delta(
(f_{\ph}^D)^{-1} - {\rm Re} {P}(\omega,\vecq)
) + R_{\rm reg}(\omega,\vecq), 
\eea 
where $R_{\rm reg}(\omega,\vecq)$ is the regular (\ie ~smooth) part of
the spectral function and $Z(\vecq)$ is the wave-function
renormalization~\cite{Keller:2012yc}. The Dirac delta gives the
spectrum of the quasiparticle excitations. In Eq.~(\ref{S_SMALL}) the
interaction term $f_{\ph}^D$ indicates that one is dealing with
density perturbations.

\begin{figure*}[t]
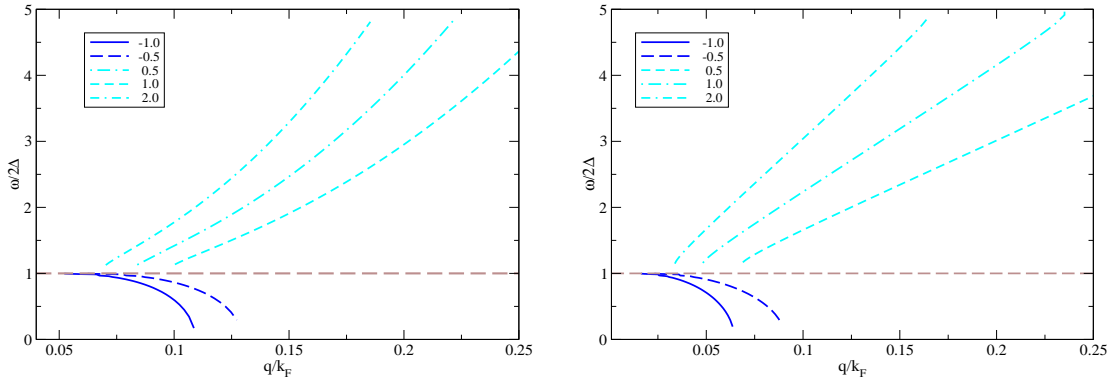

\hskip 1cm
\begin{center}
\includegraphics[width=7.0cm,height=5.0cm,angle=0]{densexitons.eps}
\hskip  0.5cm
\includegraphics[width=7.0cm,height=5.0cm,angle=0]{spinexcitions.eps}
\end{center}
\caption[] {Dependence of the frequency of the modes on 
   momentum transfer for density (left panel) and spin-density
  (right panel) perturbations~\cite{Keller:2012yc}. The values 
  of particle-hole interaction $f_{\ph}$ are shown in each panel.  
  The system parameters are fixed at $k_F = 1$ fm$^{-1}$, 
  $\Delta=1.0~\textrm{MeV}$, and $T/T_c =0.5 $.  Heavy lines 
  (blue online) correspond to undamped excitonic modes. Light 
  lines (cyan online) correspond to diffusive damped
  modes.}\label{fig_dispersion}
\end{figure*}
The dispersion relations of the excitations are determined by solutions
$\omega(\vecq)$ of the equation
\bea \label{eq:QP_spectrum} 
1-f_{\ph}^D{\rm Re}{P}(\omega,\vecq) = 0.
\eea 
Fig.~\ref{fig_dispersion} demonstrates its solutions in the case of
density and spin excitations for the range of interaction values 
$-1 \le f^D_{\ph}\le 2$. It turns out that if the interaction has a
positive sign the modes are located in the domain $\omega/\omega_0 > 1$, 
where ${\rm Im}{P}(\omega,\vecq) \neq 0$.  This is the case for 
spin excitations in the crust of a neutron star.  Such modes are 
termed {\it diffusive}.  The density modes are located in the 
domain $\omega/\omega_0 \le 1$, since the particle-hole interaction 
takes negative values in neutron-star crusts.  The pair-breaking 
part of ${\rm Im}{P}(\omega,\vecq)$ vanishes in this domain.  These 
modes can be called {\it excitonic} in that they can be viewed 
as bound pairs of particles and holes (and are analogous in this 
respect to modes occurring in ordinary semiconductors).  It must 
be noted that we cannot follow the high-momentum behavior of the 
modes because we are using perturbative response functions, which 
are no longer valid at large momentum transfers $q/k_F \sim 0.3$.  
We also observe that for negative $f_{\ph}$, the modes tend toward 
zero with increasing momentum transfer.  The high-momentum behavior 
of the excitonic modes requires further study in order to realize
a complete picture of their thermodynamics.  In the following
section we will concentrate on the properties of the spin
diffusive modes.

\section{Thermal conductivity of spin diffusive modes in neutron-star crusts}

We conclude with an estimate of the thermal conductivity of spin
diffusive modes, assuming that this property is dominated by the mode-mode
scattering process.  To treat the thermal transport in the presence of
a thermal gradient $\nabla T$, we write the kinetic equation for spin 
diffusive modes as
%--------------------------------------------------------
\bea 
\frac{\partial\varepsilon}{\partial\vecp}  \frac{\partial f_0}{\partial  T} \vecnabla T =I[f],
\eea
%--------------------------------------------------------
where $f_0$ is the equilibrium distribution of the modes, which can be
taken as a Boltzmann distribution owing to the large value of
$\omega_0$ and $I[f]$ is the collision integral.  Assuming a small
perturbation, we linearize the kinetic equation in the standard
fashion, i.e., we substitute 
%--------------------------------------------------------
\bea f = f_0+\delta f , \qquad \delta f =
-\frac{\partial f_0}{\partial \varepsilon}\chi = \frac{f_0}{T}\chi.
\eea 
%--------------------------------------------------------
in the collision integral. For this generic form of the
perturbation, the collision integral has the form 
%--------------------------------------------------------
\bea I[f_1] =
\frac{f_{01}}{T}\int w' f_{02} (\chi_1'+\chi_2'-\chi_1-\chi_2)
d\Gamma'_1 d\Gamma_2 d\Gamma'_2 , 
\eea 
%--------------------------------------------------------
where $d\Gamma_1$ etc.\ denote
the phase space volume measure and $w'$ is the transition probability. 
For thermal conductivity the perturbation is of the form 
%-------------------------------------------------------
\bea \chi = \vecg\cdot \vecnabla T = \vecv
\cdot \vecnabla T \sum_{s=1}^{\infty} A_sS^s_{3/2} (\beta\vecv), \eea
where $\vecv = ({\partial\varepsilon}/{\partial\vecp})$, $\beta =
(4\alpha T)^{-1}$, 
%--------------------------------------------------------
and the $S_r^s(x)$ are associated Laguerre
polynomials.  Truncating the expansion in polynomials by keeping only
the leading $s=1$ term, the thermal conductivity is given by $\kappa =
75/16a_{11}$ where \bea a_{11} &=&
\frac{\beta^{9/2}}{2^{5/2}\pi^{1/2}}e^{-\frac{2\omega_0}{T}} \int dv
\, v^7 e^{-\frac{\beta v^2}{2}} \sin^2\alpha \left(\frac{d\sigma
}{d\alpha}\right)d\alpha , \eea
%-------------------------------------------------
in which $d\sigma/d\alpha$ is the mode-mode scattering cross section. 
It is expected that the contribution from terms of higher order in $s$ 
is small.

We can make a simple estimate of $\kappa$ by assuming a constant 
cross section given by $d\sigma/d\Omega \simeq 1/4k_F^2$. We 
find
%-------------------------------------------------
\bea
\kappa \simeq \sqrt{\frac{2\alpha T}{\pi}}\, k_F^2 \, e^{-\frac{2\omega_0}{T}},
\eea
which exposes the temperature scaling of the thermal conductivity when the 
cross section is independent of $T$.

%-------------------------------------------------
In closing, we note the analogy of the spin diffusive modes with
rotons in superfluid helium, which can provide a guide for further
studies of scattering and transport involving these collective modes.

\section*{Acknowledgments} 
AS acknowledges support from the Deutsche Forschungsgemeinschaft
(Grant No.~SE 1836/3-1) and from the NewCompStar COST Action MP1304.
JWC acknowledges research support from the McDonnell Center for
the Space Sciences and hospitality provided by ITP, Goethe 
University, Frankfurt am Main and by the Center for Mathematical 
Sciences, University of Madeira.  This work was supported in part 
by the Helmholtz International Center for FAIR at the Goethe University,
Frankfurt am Main.

\end{document}

%% file: econfmacros-csqcd4.tex
\def\beq{\begin{equation}}
\def\eeq#1{\label{#1}\end{equation}}
\def\eeqn{\end{equation}}

\def\beqa{\begin{eqnarray}}
\def\eeqa#1{\label{#1}\end{eqnarray}}
\def\eeqan{\end{eqnarray}}

\let\bar=\overbar

\def\ie{{\it i.e.}}

\def\Dslash{\not{\hbox{\kern-4pt $D$}}}
\def\dslash{\not{\hbox{\kern-2pt $\del$}}}

\def\ee{e^+e^-}

\def\msb{{\bar{\ssstyle M \kern -1pt S}}}

\usepackage{fancyhdr,graphicx}

%% file: scr.bbl
\newpage

\begin{thebibliography}{99}
\bibitem{SedrakianClarkAlford} {\it Pairing in Fermionic Systems:
Basic Concepts and Modern Applications}, edited by 
Mark Alford, John W. Clark and Armen Sedrakian, World Scientific Publishing,
Singapore, 2006.


\bibitem{FlowersItoh1976} 


E. Flowers and N. Itoh,  ``Transport properties of dense matter,''
Astrophysical Journal {\bf 206},  218 (1976).

\bibitem{Sedrakian1996} 
A.~D.~Sedrakian, ``Neutron-Phonon Interaction in Neutron Stars: 
Phonon Spectrum of Coulomb Lattice,''
Astrophysics and Space Science {\bf 236}, 267 (1996).


\bibitem{Kobyakov:2013pza} 
 D.~Kobyakov and C.~J.~Pethick, ``Towards a metallurgy of neutron-star crusts,''
  Phys.\ Rev.\ Lett.\  {\bf 112},  112504 (2014).

 \bibitem{Pethick:2010zf} 
  C.~J.~Pethick, N.~Chamel, and S.~Reddy,
 ``Superfluid dynamics in neutron-star crusts,''
  Prog.\ Theor.\ Phys.\ Suppl.\  {\bf 186}, 9 (2010).

\bibitem{Cirigliano:2011tj} 
  V.~Cirigliano, S.~Reddy, and R.~Sharma,
  ``A low energy theory for superfluid and solid matter and 
    its application to the neutron-star crust,''
    Phys.\ Rev.\ C {\bf 84}, 045809 (2011).

\bibitem{Chamel:2012ix} 
  N.~Chamel, D.~Page and S.~Reddy,
  ``Low-energy collective excitations in the neutron-star inner 
    crust,'' Phys.\ Rev.\ C {\bf 87}, 035803 (2013).
 

\bibitem{Keller:2012yc} 
 J.~Keller and A.~Sedrakian,
``Response functions of cold neutron matter: density, 
  spin and current fluctuations,''
  Phys.\ Rev.\ C {\bf 87}, 045804 (2013).

\bibitem{Martin:2014jja} 
  N.~Martin and M.~Urban,
  ``Collective Modes in a superfluid neutron gas within the 
  Quasiparticle Random-Phase Approximation,'' 
  Phys.\ Rev.\ C {\bf 90},  065805 (2014). 
 

\bibitem{Abrikosov:1962} A. A. Abrikosov, L. P. Gor'kov, and 
I. E. Dzyaloshinski,
{\it  Methods of quantum field theory in statistical physics} 
(Dover, New York, 1975).

\bibitem{Migdal:1967}            
           A. B. Migdal, {\it Theory of Finite Fermi
           Systems and applications to Atomic Nuclei} 
           (Interscience, London, 1967).

\bibitem{Leggett:1966zz}
A.~J.~Leggett,
``Theory of a Superfluid Fermi Liquid. 2. Collective Oscillations,''
Phys.\ Rev.\  {\bf 147}, 119 (1966).




\end{thebibliography}
